%% file: double_0th.tex
\documentclass[journal]{IEEEtran}
\ifCLASSINFOpdf
\else
\fi
\ifCLASSOPTIONcompsoc
 \usepackage[caption=false,font=normalsize,labelfont=sf,textfont=sf]{subfig}
\else
 \usepackage[caption=false,font=footnotesize]{subfig}
\fi

\usepackage{graphicx}
\usepackage{amsmath}
\usepackage{mathtools}
\usepackage{epsfig}
\usepackage{float}
\usepackage{lipsum}
\usepackage{stfloats}
\usepackage{array}
\usepackage{amssymb}
\usepackage{cite}
\usepackage{url}
\usepackage{algorithm2e}
\usepackage{algorithmic}
\usepackage{multirow}
\usepackage{fancyh}
\usepackage{upgreek}
\usepackage{dsfont}
\usepackage{gensymb}

\usepackage{multicol, blindtext}
\usepackage{mathtools, cuted}

\usepackage{color}
\usepackage{bm}
\usepackage{booktabs}
\usepackage{tablefootnote}
\usepackage{threeparttable}

\usepackage{multicol, blindtext}
\usepackage{mathtools, cuted}

\usepackage{color}
\usepackage{bm}
\usepackage{booktabs}
\usepackage{tablefootnote}
\usepackage{threeparttable}

\DeclareMathOperator*{\argmax}{argmax} % no space, limits underneath in displays

\normalsize
\graphicspath{ {./images/} }

\input{input.tex}
\begin{document}
\title{Finding Globally Optimal Configuration of\\Active RIS in Linear Time 
\thanks{}
}

\author{\IEEEauthorblockN{Heedong~Do},
 {\it Member,~IEEE},
\and
\IEEEauthorblockN{Namyoon~Lee},
{\it Senior Member,~IEEE},
% \and\\
% \IEEEauthorblockN{Angel~Lozano},
% {\it Fellow,~IEEE}
}
\maketitle

\maketitle

\begin{abstract}
This paper presents an algorithm for finding the optimal configuration of active reconfigurable intelligent surface (RIS) when both transmitter and receiver are equipped with a single antenna each. The resultant configuration is globally optimal and it takes linear time for the computation. Moreover, there is a closed-form expression for the optimal configuration when the direct link vanishes, which enables further analysis.
\end{abstract}



% \fi
\IEEEpeerreviewmaketitle

\section{Introduction}

Of late, nonregenerative relay technologies have revived with new flavors and brand-new names---reconfigurable intelligent surfaces (RISs) or intelligent reflecting surfaces \cite{liaskos2018new, wu2019towards, basar2019wireless,do2022line}. The most contrasting features compared to their predecessor are that they lack radio-frequency chains and baseband processing, which potentially make them more economical than their competitors.

The primary objective of the RISs is, of course, to extend a coverage by circumventing the obstacles. To do so, RISs should not be placed in the vicinity of the base station to maximize their coverage extension capability. When it comes to passive RISs which merely reradiate the impinging signal with phase shifts, this conflicts with the placement guideline, placing the RIS near either end \cite{dunna2020scattermimo, wu2021intelligent} to maximize a receive SNR. This fundamental trade-off can be lifted by the use of active RISs, which amplify and then reradiate the impinging signal \cite{abari2017enabling, you2021wireless, long2021active, zhi2022active, zhang2022active, shu2023three, lin2023enhanced}, corroborated by experimental studies \cite[Table II]{rao2023active}.

To play their role as premised, RISs should be configured properly. For passive RISs, with the achievable rate as an objective, phase shift optimization has been studied in a variety of contexts, to wit: single-input single-output (SISO) \cite{bjornson2019intelligent}, multiple-output single-input (MISO)/single-input multiple-output (SIMO) \cite{yu2019miso}, multiple-input multiple-output (MIMO) \cite{zhang2020capacity, perovic2021achievable}, and multi-user MISO \cite{wu2019intelligent, guo2020weighted, pan2020multicell, xie2020max, liu2021two}.
% Some underlying structures of channels could greatly facilitate the optimization: sparse channels \cite{wang2020intelligent} and line-of-sight MIMO channels \cite{do2022line, bartoli2022spatial}.
When it comes to active RISs, analogous problems have been tackled for SISO \cite{shu2023three, lin2023enhanced}, MISO/SIMO \cite{long2021active}, and multi-user MISO \cite{zhang2022active} systems.

Unlike the passive counterpart, where a trivial closed-form solution is available \cite{bjornson2019intelligent}, the SISO problem with active RIS is still not satisfactorily solved. Although several iterative algorithms have been devised \cite{long2021active, zhang2022active, shu2023three, lin2023enhanced}, there is no guarantee of global optimality owing to the non-convex nature of the problem (see Sec. \ref{optimalIrsPhases}). That said, it is not always true that global optimization of a non-convex function is computationally infeasible; recall that even the optimization problem for SISO case with passive RISs technically is non-convex. 
% (see more examples of tractable non-convex problems in \cite{kha2011fast, do2020reconfigurable, ren2022linear}).
To this end, we scrutinize the simplest SISO case with active RIS in the present paper.

The key findings of the present paper are
\begin{itemize}
    \item When the direct link is blocked, the \textit{globally optimal} configuration can be computed in a closed-form, whose computational complexity is \textit{linear} in the number of RIS elements. 
    \item In the presence of the direct link, the \textit{globally optimal} configuration can be found by one-dimensional root finding whose per-iteration complexity is \textit{linear}.
\end{itemize}
Given that the state-of-the-art methods are not guaranteed to be globally optimal and have cubic complexity at least, our result greatly extends the existing results.
% Particularly, we develop an optimization algorithm, and
% \begin{itemize}
%     \item The algorithm is \textit{globally optimal}.
%     \item Its complexity is \textit{linear} in the number of RIS elements, whereas the existing suboptimal algorithms are of cubic order at least.
% \end{itemize}
% Moreover, when the direct link is blocked, the optimal configuration is given even in closed-form.

Albeit concerning the simplest SISO setup, to the best of our knowledge, it is the first concrete nontrivial example of global optimization for active RISs. It could also serve as a useful benchmark for algorithms developed in more general settings. Moreover, it can be used as a subroutine in the MISO/SIMO system \cite{long2021active}.

The paper is organized as follows. Sec. \ref{systemModel} introduces the system model, and a comparison with amplify-and-forward (AF) relays is made in Sec. \ref{comparisonWithAfRelays}. The optimal RIS phases are described in Sec. \ref{optimalIrsPhases}. With the optimal RIS phases, the problem boils down to the amplitude-only problem whose properties are investigated in Sec. \ref{takingCloserLookAtOptimizationProblem}. The methods for computing the optimal amplitudes are presented in Secs. \ref{whenDirectLinkDoesNotExist} and \ref{whenDirectLinkExists} with a numerical validation in Sec. \ref{assessingExistingAlgorithms}. Tractable hardware constraints are discussed in Sec. \ref{incormporatingAdditionalRisConstraints}. Finally, the paper concludes in Sec. \ref{summary} with a summary.

\section{System Model}
\label{systemModel}
Consider a single-antenna transmitter, a single-antenna receiver, and an $N$-element RIS. Denoting the transmit signal by $s\sim \cN_{\bbC}(0,P_1)$, the receive signal $y$ can be modelled as
\begin{align}
    y &= \underbrace{h_{\rm d}s}_{\mathclap{\text{Direct link}}}+\underbrace{\bh_2^\top \bPhi(\bh_1 s+\bv_1)}_{\mathclap{\text{RIS-assisted link}}}+v_2\\
    &= (h_{\rm d}+\bh_2^\top\bPhi \bh_1)s + \underbrace{\bh_2^\top \bPhi \bv_1 + v_2}_{\mathclap{\text{Effective noise}}}. \label{signalModel}
\end{align}
Here, $h_{\rm d}\in \bbC$, $\bh_1\in\bbC^{N\times 1}$, and $\bh_2\in\bbC^{N\times 1}$ denote the channel from the transmitter to the receiver, from the transmitter to the RIS, and from the RIS to the receiver, respectively; $\bPhi\in\bbC^{N\times N}$
denotes the RIS phase shifts, which is diagonal; and $\bv_1 \sim \cN_{\bbC}(\boldsymbol{0},\sigma_1^2\bI)$ and $v_2 \sim \cN_{\bbC}(0,\sigma_2^2)$ are Gaussian noise at the RIS and the receiver, respectively.
With a power constraint of $P_2$ at the RIS, we have
\begin{align}
    &\bbE[\|\bPhi(\bh_1 s+\bv_1)\|^2]
    = P_1\|\bPhi \bh_1\|^2  + \|\bPhi\|_{\rm F}^2 \leq P_2. \label{powerConstraint}
    % \nonumber\\
    % &\qquad\qquad\qquad = P_1\sum_n |[\bh_1]_n|^2  p_n^2 + \sigma_1^2\sum_n p_n^2 \leq P_2,
\end{align}
Here, $\|\cdot\|_{\rm F}$ denotes the Frobenius norm.

Throughout the paper, we assume that all channels are constant and the channel state information (CSI) at the RIS is perfect.

\subsection{Useful Normalization}
We can rewrite the signal model \eqref{signalModel} and the power constraint \eqref{powerConstraint} as follows:
\begin{align}
    \frac{y}{\sigma_2} &= \bigg(\frac{\sigma_1}{\sigma_2}h_{\rm d}+\bh_2^\top\bigg(\frac{\sigma_1}{\sigma_2} \bPhi\bigg) \bh_1\bigg) \frac{s}{\sigma_1} \nonumber\\
    &\qquad\qquad\qquad\qquad + \bh_2^\top \bigg(\frac{\sigma_1}{\sigma_2} \bPhi\bigg) \frac{\bv_1}{\sigma_1} + \frac{v_2}{\sigma_2}\nonumber\\
    & \frac{P_1}{\sigma_1^2}\bigg\|\frac{\sigma_1}{\sigma_2}\bPhi \cdot \bh_1\bigg\|^2  + \bigg\|\frac{\sigma_1}{\sigma_2}\bPhi\bigg\|_{\rm F}^2 \leq \frac{P_2}{\sigma_2^2}
\end{align}
The substitutions
\begin{align}
    &s \gets \frac{s}{\sigma_1} && y \gets \frac{y}{\sigma_2} && h_{\rm d} \gets \frac{\sigma_1}{\sigma_2} h_{\rm d} \nonumber\\
    &\bPhi \gets \frac{\sigma_1}{\sigma_2} && \bv_1 \gets \frac{\bv_1}{\sigma_1} && v_2 \gets \frac{v_2}{\sigma_2},
\end{align}
in part borrowed from \cite[Eqn. 12]{tang2007optimal}, greatly streamline the signal model; we still have \eqref{signalModel} but with the signals
\begin{align}
    s \sim \cN_\bbC(0,\SNR_1) \quad\,\,\,\bv_1 \sim \cN_\bbC(\boldsymbol{0},\bI) \quad\,\,\, v_2 \sim \cN_\bbC(0,1).
\end{align}
With the substitution, the power constraint becomes
\begin{align}
    \SNR_1\|\bPhi \bh_1\|^2  + \|\bPhi\|_{\rm F}^2 \leq \SNR_2, \label{powerConstraintSimplified}
\end{align}
where
\begin{align}
    \SNR_1 = \frac{P_1}{\sigma_1^2} \qquad \SNR_2 = \frac{P_2}{\sigma_2^2}. \label{snrDefinition}
\end{align}

We hereafter work with this simplified model.
The receive SNR is then
\begin{align}
    &\frac{\bbE\big[|(h_{\rm d}+\bh_2^\top\bPhi \bh_1)s|^2\big]}{\bbE\big[|\bh_2^\top \bPhi \bv_1 + v_2|^2\big]} 
     = \frac{|h_{\rm d}+\bh_2^\top\bPhi \bh_1|^2}{1+\|\bh_2^\top \bPhi\|^2}\SNR_1. \label{receivedSnrActiveIrs}
    %  \nonumber\\
    % &\qquad\qquad\qquad = \frac{|h_{\rm d}+\sum_n [\bh_1]_n [\bh_2]_n (p_n e^{j\phi_n})|^2 P_1}{\sigma_1^2\sum_n |[\bh_2]_n|^2p_n^2 + \sigma_2^2}.
\end{align}
With the power budget \eqref{powerConstraintSimplified},
the optimization of the receive SNR is then
% \begin{align}
%     \max_{\{p_n\},\{\phi_n\}} \quad& \frac{|h_{\rm d}+\sum_n [\bh_1]_n [\bh_2]_n (p_n e^{j\phi_n})|^2 P_1}{\sigma_1^2\sum_n |[\bh_2]_n|^2p_n^2 + \sigma_2^2}\\
%     \text{s.t.} \quad& P_1\sum_n |[\bh_1]_n|^2  p_n^2 + \sigma_1^2\sum_n p_n^2 \leq P_2. \nonumber
% \end{align}
\begin{align}
    \max_{\bPhi} \quad& \frac{|h_{\rm d}+\bh_2^\top\bPhi \bh_1|^2}{1+\|\bh_2^\top \bPhi\|^2} \SNR_1\\
    \text{s.t.} \quad& \SNR_1\|\bPhi \bh_1\|^2  + \|\bPhi\|_{\rm F}^2 \leq \SNR_2, \nonumber\\
    & \bPhi=\diag(p_1e^{j\phi_1},\ldots,p_Ne^{j\phi_N}). \nonumber
\end{align}

\subsection{Power Constraint at the Transmitter}
We have implicitly assumed $\bbE[|s|^2] = \SNR_1$ in lieu of $\bbE[|s|^2] \leq \SNR_1$. That said, this additional constraint does not incur any loss of generality. Put differently, we need to use all the allowed power at the transmitter to maximize the receive SNR.

To verify it, it is sufficient to show that \eqref{receivedSnrActiveIrs} is increasing with respect to $\SNR_1$.
Let $\SNR_1 < \SNR_1'$ and denote the optimal phase shifts for $\SNR_1$ by $\bPhi$. It can be shown by showing a suboptimal choice
\begin{align}
    \bPhi' = \sqrt{\frac{\SNR_1}{\SNR_1'}}\bPhi. \label{suboptimalChoice}
\end{align}
at $\SNR_1'$ outperforms $\bPhi$ at $\SNR_1$:
\begin{align}
    \frac{|h_{\rm d}+\bh_2^\top\bPhi' \bh_1|^2}{1+\|\bh_2^\top \bPhi'\|^2} \SNR_1' & = \frac{\Big|\sqrt{\frac{\SNR_1'}{\SNR_1}}h_{\rm d}+\bh_2^\top\bPhi \bh_1\Big|^2}{1+\Big\|\sqrt{\frac{\SNR_1}{\SNR_1'}}\bh_2^\top \bPhi\Big\|^2} \SNR_1\nonumber\\
    &\geq \frac{|h_{\rm d}+\bh_2^\top\bPhi \bh_1|^2}{1+\|\bh_2^\top \bPhi\|^2} \SNR_1.
\end{align}
The choice \eqref{suboptimalChoice} abides by the power constraint since
\begin{align}
    &\SNR_1'\|\bPhi' \bh_1\|^2  + \|\bPhi'\|_{\rm F}^2 \\
    &=\SNR_1\|\bPhi \bh\|^2 + \frac{\SNR_1}{\SNR_1'}\|\bPhi\|_{\rm F}^2 \\
    &=\SNR_1\|\bPhi \bh\|^2 + \|\bPhi\|_{\rm F}^2 \leq \SNR_2.
\end{align}

\section{Comparison with AF Relays}
\label{comparisonWithAfRelays}
From a mathematical standpoint, the system model rather resembles that of AF relays with time division multiple access. With the same substitutions above, we have the following signal model \cite{fang2006joint, tang2007optimal, taricco2022information}:
\begin{align}
    \by
    = 
    \begin{bmatrix}
    h_{\rm d} \\ \bh_2^\top\bPhi\bh_1
    \end{bmatrix}
    s  
    + \begin{bmatrix}
    v_0 \\ \bh_2^\top\bPhi\bv_1+v_2
    \end{bmatrix},
\end{align}
where the signals are stacked vertically in time and $v_0\sim \cN_{\bbC}(0,1)$ is the noise at the first slot. 

After normalizing the noise of the second slot, that is, 
\begin{align}
    \begin{bmatrix}
    1 & 0\\ 0 & \frac{1}{\sqrt{1+\|\bh_2^\top \bPhi\|^2}}
    \end{bmatrix}
    \by
    = 
    \begin{bmatrix}
    h_{\rm d} \\ \frac{\bh_2^\top\bPhi\bh_1}{\sqrt{1+\|\bh_2^\top \bPhi\|^2}}
    \end{bmatrix}
    s  
    + \begin{bmatrix}
    v_0 \\ \frac{\bh_2^\top\bPhi\bv_1+v_2}{\sqrt{1+\|\bh_2^\top \bPhi\|^2}}
    \end{bmatrix}.
\end{align}
This is the signal model for single-input multi-output system and the optimal processing at the receiver is maximum ratio combining. Hence, the receive SNR is
\begin{align}
    \bigg(|h_{\rm d}|^2 + \frac{|\bh_2^\top\bPhi\bh_1|^2}{1+\|\bh_2^\top \bPhi\|^2}\bigg)\SNR_1
\end{align}
As the first term is constant, the term
\begin{align}
    \frac{|\bh_2^\top\bPhi\bh_1|^2}{1+\|\bh_2^\top \bPhi\|^2}\SNR_1 \label{receivedSnrRelay}
\end{align}
is of importance, which is identical to the receive SNR \eqref{receivedSnrActiveIrs} for active RIS in the absence of direct path. The only difference is that there is no diagonal constraint for AF relays, resulting in the following optimization problem:
\begin{align}
    \max_{\bPhi} \quad& \frac{|\bh_2^\top\bPhi\bh_1|^2}{1+\|\bh_2^\top \bPhi\|^2}\SNR_1 \\
    \text{s.t.} \quad& \SNR_1\|\bPhi \bh_1\|^2  + \|\bPhi\|_{\rm F}^2 \leq \SNR_2. \nonumber
\end{align}
The very problem is essentially a special case of the problem addressed in \cite{fang2006joint, tang2007optimal, taricco2022information}, and the maximum of \eqref{receivedSnrRelay} is
\begin{align}
    \frac{\|\bh_1\|^2\|\bh_2\|^2}{\|\bh_1\|^2\SNR_1 + \|\bh_2\|^2\SNR_2+1} \SNR_1\SNR_2. \label{receivedSnrMaximum}
\end{align}
Having said that, we provide a simple elementary proof in App. \ref{relayProof} for clarity.

\section{Optimal RIS Phases}
\label{optimalIrsPhases}
From triangle inequality, it is apparent that we should set
\begin{align}
    \phi_n = \angle h_{\rm d} - \angle [\bh_2]_n -\angle [\bh_1]_n,
\end{align}
where $\angle\,\cdot$ denotes the argument of a complex number.
With $\{p_n\}$ held fixed, it maximizes the signal power while keeping the noise power, which is recognized in \cite{long2021active}.

Let us embrace the substitutions
\begin{align}
    h_{\rm d} \gets |h_{\rm d}| \qquad [\bh_2]_n \gets |[\bh_2]_n| \qquad [\bh_1]_n \gets |[\bh_1]_n|
\end{align}
for brevity.
The problem, in turn, boils down to
\begin{align}
    \max_{\{p_n\}} \quad& \frac{(h_{\rm d}+\sum_n [\bh_1]_n [\bh_2]_n p_n )^2}{1 + \sum_n [\bh_2]_n^2p_n^2}\SNR_1 \\
    \text{s.t.} \quad& \sum_n ([\bh_1]_n^2\SNR_1+1) p_n^2 \leq \SNR_2. \nonumber
\end{align}
Even without the constraint $p_n\geq 0$ for all $n$, the optimal amplitudes $\{p_n\}$ of \eqref{simplifiedProblem} are positive from the positivity of $h_{\rm d}$. In this regard, we need not append an additional constraint.

Introducing new variables,
\begin{align}
    \alpha_n &= [\bh_1]_n[\bh_2]_n\\
    \beta_n &= [\bh_2]_n^2\\
    \gamma_n &= \frac{[\bh_1]_n^2\SNR_1+1}{\SNR_2},
\end{align}
and omitting the irrelevant constant $\SNR_1$, the optimization problem can be recast as
\begin{align}
    \max_{\{p_n\}} \quad&  \frac{(h_{\rm d} + \sum_n \alpha_n p_n)^2}{1+ \sum_n \beta_n p_n^2} \label{simplifiedProblem}
    % 2\log\bigg(1+\sum_n \alpha_n p_n\bigg)-\log\bigg(1+ \sum_n \beta_n p_n^2\bigg) 
    \\
    \text{s.t.} \quad& \sum_n \gamma_n p_n^2 -1\leq 0. \nonumber
\end{align}

\begin{figure}
    \centering
    \includegraphics[width = 0.9\linewidth]{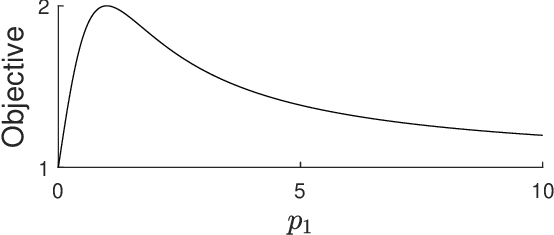}
    \caption{Restriction \eqref{objectiveOnRay} of the objective function to the ray with $h_{\rm d} = \alpha_1 = \beta_1 = 1$.}
    \label{non-convexity}
\end{figure}

\section{Taking Closer Look at Optimization Problem}
\label{takingCloserLookAtOptimizationProblem}
This section touches on the properties of the optimization problem \eqref{simplifiedProblem}.
Both good and bad news are put forth.
 
\subsection{Non-Convexity}
In general, the objective of \eqref{simplifiedProblem} is unfortunately not concave. To show this, it is sufficient to show that the restriction of the objective to the ray $[0,\infty)\times \{0\} \times \ldots \times \{0\}$,
\begin{align}
    \frac{(h_{\rm d} + \alpha_1 p_1)^2}{1+ \beta_1 p_1^2}, \label{objectiveOnRay}
\end{align}
is not concave (see Fig. \ref{non-convexity}), which can be easily verified using elementary calculus. This precludes the use of off-the-shelf solvers and has motivated alternative approaches.

\subsection{Unique Stationary Point}
\label{uniqueStationaryPoint}
Albeit being non-convex, the objective function has a unique stationary point when $h_{\rm d} \neq 0$. 
This result is presented in \cite[Prop. 1]{long2021active}, however the detailed derivation is omitted. For the sake of completeness, we present the derivation since it take only a few lines to derive.

Taking logarithm and differentiating with respect to $p_k$, we obtain
\begin{align}
    \frac{2\alpha_k}{h_{\rm d}+\sum_n \alpha_n p_n}-\frac{2\beta_k p_k}{1+\sum_n \beta_n p_n^2}.
\end{align}
Equating it with zero gives
\begin{align}
    p_k = \frac{\alpha_k}{\beta_k}\rho \label{amplitudeProportionalToFraction}
\end{align}
where
\begin{align}
    \rho = \frac{1+\sum_n \beta_n p_n^2}{h_{\rm d}+\sum_n \alpha_n p_n}. \label{multiplicativeConstant}
\end{align}
Plugging \eqref{amplitudeProportionalToFraction} into \eqref{multiplicativeConstant}, we have
\begin{align}
    \rho=\frac{1+\sum_n \frac{\alpha_n^2}{\beta_n}\rho^2}{h_{\rm d}+\sum_n \frac{\alpha_n^2}{\beta_n}\rho}.
\end{align}
Simplifying it, we obtain $\rho=\frac{1}{h_{\rm d}}$. That is, the unique stationary point is
% \begin{align}
%     (p_1,\ldots,p_N) &= \bigg(\frac{\alpha_1}{h_{\rm d} \beta_1},\ldots,\frac{\alpha_N}{h_{\rm d} \beta_N}\bigg)\\
%     &=\bigg(\frac{[\bh_1]_1}{h_{\rm d} [\bh_2]_1},\ldots,\frac{[\bh_1]_N}{h_{\rm d} [\bh_2]_N}\bigg).
% \end{align}
\begin{align}
    p_n = \frac{\alpha_n}{h_{\rm d} \beta_n}=\frac{[\bh_1]_n}{h_{\rm d} [\bh_2]_n}.
\end{align}
It outputs the objective
\begin{align}
    \frac{\big(h_{\rm d} + \frac{1}{h_{\rm d}}\sum_n \frac{\alpha_n^2}{\beta_n}\big)^2}{1+ \frac{1}{h_{\rm d}^2}\sum_n \frac{\alpha_n^2}{\beta_n}} &= h_{\rm d}^2 + \sum_n \frac{\alpha_n^2}{\beta_n}\\
    &=h_{\rm d}^2 + \|\bh_1\|^2. \label{optimalObjectiveInfiniteBudget}
\end{align}
% and the corresponding objective is
% \begin{align}
%     1+\sum_n \frac{\alpha_n^2}{\beta_n} = 1+\frac{\|\bh_1\|^2}{h_{\rm d}^2}. \label{optimalObjectiveInfiniteBudget}
% \end{align}
%  It is interesting that it is identical to the SNR gain when regarding the RIS and the receiver as a single entity.

It is further claimed in \cite[Prop. 1]{long2021active} that it is the global maximum of unconstrained version of \eqref{simplifiedProblem}. That said, unique local extremum is not guaranteed to be global extremum (see \cite{rosenholtz1985only} for counterexample). 
% To fill the tiny gap, we prepare App. \ref{amplitudeWithoutConstraintProof}.
To fill the tiny gap, we prepare an alternative proof.
By means of Cauchy-Schwarz inequality, we have
% \begin{align}
%     &\bigg(\!1+\sum_n [\bh_2]_n^2p_n^2\!\bigg)\bigg(\!h_{\rm d}^2+\sum_n [\bh_1]_n^2\!\bigg) \nonumber\\
%     &\qquad\qquad\qquad\qquad \geq \bigg(\!h_{\rm d}+\sum_n [\bh_1]_n [\bh_2]_n p_n \!\bigg)^2
% \end{align}
\begin{align}
    \bigg(1\!+\!\sum_n \beta_n p_n^2\bigg)\!\bigg(h_{\rm d}^2\!+\!\sum_n \frac{\alpha_n^2}{\beta_n}\bigg) \!\geq\! \bigg(h_{\rm d}\!+\!\sum_n \alpha_n p_n \bigg)^2
\end{align}
whose equality condition is 
\begin{align}
    p_k &= \frac{\alpha_k}{\beta_k}=\frac{[\bh_1]_k}{h_{\rm d}[\bh_2]_k}. \label{optimalPowerInfiniteBudget}
\end{align}

The unconstrained setup corresponds to the limiting case $\SNR_2\rightarrow \infty$ where the power constraint at the RIS becomes obsolete.
The chief conclusion is that it is better to use finite power at the 
RIS even if there is no limit. This is well-known fact for AF relays \cite[Sec. 3.5]{bharadia2014fastforward} (see also \cite{zelaya2021lava}).

Let us get back to the original problem. Provided that
\begin{align}
    &\sum_n \frac{\alpha_n^2 \gamma_n}{\beta_n^2} \leq h_{\rm d}^2
    % &\qquad \Leftrightarrow h_{\rm d}^2\SNR_2 \geq \bigg(\sum_n \frac{[\bh_1]_n^4}{[\bh_2]_n^2}\bigg) \SNR_1 + \sum_n \frac{[\bh_1]_n^2}{[\bh_2]_n^2},
    \label{highSnrCondition}
\end{align}
we can attain the objective \eqref{optimalObjectiveInfiniteBudget}, which cannot be further improved. Put another way, this result characterizes the optimal $\{p_n\}$ when \eqref{highSnrCondition} holds.

Let us consider the remaining case, i.e., $h_d=0$. Cauchy-Schwarz inequality gives
\begin{align}
    \frac{(\sum_n \alpha_n p_n)^2}{1+ \sum_n \beta_n p_n^2} < \frac{(\sum_n \alpha_n p_n)^2}{\sum_n \beta_n p_n^2} \leq \sum_n\frac{\alpha_n^2}{\beta_n}. \label{noDirectLinkInfiniteBudget}
\end{align}
One can approach the bound by letting $p_k = \frac{\alpha_k}{\beta_k}\rho$ with $\rho\rightarrow \infty$:
\begin{align}
    \frac{(\sum_n \alpha_n p_n)^2}{1+ \sum_n \beta_n p_n^2} \leq \frac{\big(\sum_n \frac{\alpha_n^2}{\beta_n}\big)^2\rho^2}{1+ \big(\sum_n \frac{\alpha_n^2}{\beta_n}\big)\rho^2}
    \rightarrow \sum_n\frac{\alpha_n^2}{\beta_n}.
\end{align}
Conversely, we cannot attain the bound with finite power owing to the strict inequality in \eqref{noDirectLinkInfiniteBudget}.

\subsection{Finding the Global Optimum by Taking Square Root}
Taking a square root to the objective, the problem \eqref{simplifiedProblem} can be alternatively written as
\begin{align}
    \max_{\{p_n\}}\quad& \frac{h_{\rm d}+\sum_n \alpha_n p_n}{\sqrt{1+\sum_n \beta_n p_n^2}} \label{fractionalProgramming}\\
    \text{s.t.} \quad& \sum_n \gamma_n p_n^2 -1\leq 0, \nonumber
\end{align}
where the numerator and the denominator of the objective are concave and convex, respectively. Taking the square root does the trick here; without this, we cannot make the objective in a desired form---concave over convex.
Note that the convexity of the denominator can be derived straightforwardly as it is a composition of affine and convex mappings:
\begin{align}
    (p_1,\ldots,p_N) &\mapsto (1,\sqrt{\beta_1}p_1,\ldots,\sqrt{\beta_N}p_N) \nonumber\\
    &\mapsto \|(1,\sqrt{\beta_1}p_1,\ldots,\sqrt{\beta_N}p_N)\|.
\end{align}

This class of problem is termed concave-convex fractional programming (see \cite{shen2018fractional} for its applications for communication systems). Applying Dinkelbach's transform, it can be transformed into a series of convex problems whose solution converges to the global optimum  \cite{dinkelbach1967nonlinear}.
% \cite{dinkelbach1967nonlinear, shen2018fractional, shen2018fractional2}.
As shown in App. \ref{fractionalProof}, each problem is in fact second-order cone programming (SOCP), which can be solved using off-the-shelf solvers such as CVX \cite{cvx}.
This simple fact has not been acknowledged previously.

In the subsequent sections, we will present another globally optimal algorithms, which are more computationally efficient than the approach above.
% In \cite{lin2023enhanced}, an algorithm inspired by fractional programming is proposed to solve the very problem. That said, this approach does not have a guarantee of global optimality.

% The complexity of solving a series of convex problems obtained from Dinkelbach's transform is $\cO()$ (see App. \ref{fractionalProof}), which still has room for improvement.
% \heedong{PENDING...}

\section{Optimal Amplitudes When Direct Link\\Does Not Exist}
\label{whenDirectLinkDoesNotExist}

This section presents a closed-form solution for \eqref{simplifiedProblem} when $h_{\rm d}=0$.
Albeit being a non-convex problem, we can find the global optimum which is even in closed-form. Replacing one in the denominator of the objective with $\gamma_n p_n^2$ does the trick:
\begin{align}
    \frac{(\sum_n \alpha_n p_n)^2}{1+ \sum_n \beta_n p_n^2} &\leq  \frac{(\sum_n \alpha_n p_n)^2}{\sum_n (\beta_n+\gamma_n) p_n^2}\\
    &\leq \sum_n \frac{\alpha_n^2}{\beta_n+\gamma_n},
\end{align}
where the last step follows from Cauchy-Schwarz inequality. The equality conditions are
\begin{align}
    \sum_n \gamma_n p_n^2 -1 = 0 \qquad\quad p_n= \frac{\alpha_n}{\beta_n+\gamma_n}\cdot {\sf constant}. 
\end{align}
These conditions can be simultaneously fulfilled by
\begin{align}
    p_k &= \frac{\alpha_k}{\beta_k+\gamma_k}\bigg(\sum_n \frac{\alpha_n^2\gamma_n}{(\beta_n+\gamma_n)^2}\bigg)^{-\frac{1}{2}}, \label{closedFormAmplitude}
\end{align}
which can be expanded as
\begin{align}
    &p_k = \frac{[\bh_1]_k[\bh_2]_k}{[\bh_1]_k^2\SNR_1+[\bh_2]_k^2\SNR_2+1} \label{closedFormAmplitudeExpanded}\\
    &\!\cdot\! \bigg(\! \sum_n \frac{[\bh_1]_n^2\SNR_1\!+\!1}{\SNR_2}\!\cdot\! \frac{[\bh_2]_n^2[\bh_1]_n^2}{([\bh_1]_n^2\SNR_1\!+\![\bh_2]_n^2\SNR_2\!+\!1)^2}\!\bigg)^{\!-\frac{1}{2}}. \nonumber
\end{align}
% \begin{figure*}
% \begin{align}
%     p_k = \frac{[\bh_1]_k[\bh_2]_k}{[\bh_1]_k^2\SNR_1+[\bh_2]_k^2\SNR_2+1} \cdot \bigg(\! \sum_n \frac{[\bh_1]_n^2\SNR_1+1}{\SNR_2}\cdot \frac{[\bh_2]_n^2[\bh_1]_n^2}{([\bh_1]_n^2\SNR_1+[\bh_2]_n^2\SNR_2+1)^2}\!\bigg)^{\!-\frac{1}{2}}. \label{closedFormAmplitudeExpanded}
% \end{align}
% \hrulefill
% \end{figure*}
The maximum receive SNR is therefore
\begin{align}
    &\frac{ \big(\sum_n \frac{\alpha_n^2}{\beta_n+\gamma_n}\big)^2 / \big(\sum_n \frac{\alpha_n^2\gamma_n}{(\beta_n+\gamma_n)^2}\big)}{1+ \big(\sum_n \frac{\alpha_n^2 \beta_n}{(\beta_n+\gamma_n)^2}\big) / \big(\sum_n \frac{\alpha_n^2\gamma_n}{(\beta_n+\gamma_n)^2}\big) }\SNR_1\\
    &=\bigg(\sum_n \frac{\alpha_n^2}{\beta_n+\gamma_n}\bigg)\SNR_1\\
    &= \bigg(\sum_n \frac{[\bh_2]_n^2 [\bh_1]_n^2}{[\bh_1]_n^2 \SNR_1+[\bh_2]_n^2\SNR_2+1}\bigg) \SNR_1 \SNR_2. \label{receivedSnrIrsMaximum}
\end{align}

The computational complexity $\{p_n\}$ in \eqref{closedFormAmplitude} is $\cO(N)$ as it is sufficient to compute the normalizing constant $\sum_n \frac{\alpha_n^2\gamma_n}{(\beta_n+\gamma_n)^2}$
only once and reuse it.

% Let us start from \eqref{receivedSnrActiveIrs}. We immediately have
% \begin{align}
%     &\frac{(\sum_n [\bh_1]_n [\bh_2]_n p_n )^2}{1+\sum_n [\bh_2]_n^2p_n^2 } \SNR_1\\
%     &=\frac{(\sum_n [\bh_1]_n [\bh_2]_n p_n )^2}{\SNR_2+\sum_n [\bh_2]_n^2\SNR_2 p_n^2 } \SNR_1 \SNR_2\\
%     &\leq \frac{(\sum_n [\bh_1]_n [\bh_2]_n p_n )^2}{\sum_n ([\bh_1]_n^2 \SNR_1+[\bh_2]_n^2\SNR_2+1) p_n^2 } \SNR_1 \SNR_2\\
%     &\leq \bigg(\sum_n \frac{[\bh_2]_n^2 [\bh_1]_n^2}{[\bh_1]_n^2 \SNR_1+[\bh_2]_n^2\SNR_2+1}\bigg) \SNR_1 \SNR_2
% \end{align}
% where the first inequality follows from the power constraint
% while the second inequality follows from the Cauchy-Schwarz inequality. The equality conditions are
% \begin{align}
%     & \sum_n ([\bh_1]_n^2 \SNR_1+1) p_n^2 = \SNR_2\\
%     & p_n \propto \frac{[\bh_1]_n [\bh_2]_n}{[\bh_1]_n^2 \SNR_1+[\bh_2]_n^2\SNR_2+1},
% \end{align}
% respectively. These conditions can be simultaneously fulfilled by 
% \begin{align}
%     p_k &= \frac{[\bh_1]_k[\bh_2]_k}{[\bh_1]_k^2\SNR_1+[\bh_2]_k^2\SNR_2+1} \cdot \bigg(\!\frac{[\bh_1]_n^2\SNR_1+1}{\SNR_2} \nonumber\\
%     &\qquad \cdot\sum_n \frac{[\bh_2]_n^2[\bh_1]_n^2}{([\bh_1]_n^2\SNR_1+[\bh_2]_n^2\SNR_2+1)^2}\!\bigg)^{\!-\frac{1}{2}}.
% \end{align}

\begin{table*}[]
\centering
\begin{threeparttable}
\setlength{\tabcolsep}{5pt}
\caption{Comparison of Receive SNR\tnote{*}}
\label{comparisonOfReceivedSnr}
\begin{tabular}{cccc} 
\toprule
 & Active RIS (equal amplitudes) & Active RIS (optimal amplitudes) & AF relay\\
\midrule
General  & $\frac{(\sum_n [\bh_1]_n [\bh_2]_n)^2}{\|\bh_1\|^2 \SNR_1+\|\bh_2\|^2\SNR_2+N} \SNR_1 \SNR_2 $ & $\Big(\sum_n \frac{[\bh_1]_n^2 [\bh_2]_n^2}{[\bh_1]_n^2 \SNR_1+[\bh_2]_n^2\SNR_2+1}\Big)\SNR_1 \SNR_2$ & $\frac{\|\bh_1\|^2\|\bh_2\|^2}{\|\bh_1\|^2\SNR_1 + \|\bh_2\|^2\SNR_2+1}\SNR_1\SNR_2$\\
$\SNR_1\rightarrow 0^+$ & $\frac{(\sum_n [\bh_1]_n [\bh_2]_n)^2\SNR_2}{\|\bh_2\|^2\SNR_2+N} \SNR_1 + \cO(\SNR_1^2) $ & $\Big(\sum_n \frac{[\bh_1]_n^2 [\bh_2]_n^2\SNR_2}{[\bh_2]_n^2\SNR_2+1}\Big)\SNR_1 + \cO(\SNR_1^2) $ & $\frac{\|\bh_1\|^2\|\bh_2\|^2\SNR_2}{\|\bh_2\|^2\SNR_2+1}\SNR_1 + \cO(\SNR_1^2)$\\
$\SNR_1\rightarrow \infty$ & $\frac{(\sum_n [\bh_1]_n [\bh_2]_n)^2}{\|\bh_1\|^2} \SNR_2 + \cO(\SNR_1^{-1}) $ & $\|\bh_2\|^2 \SNR_2 + \cO(\SNR_1^{-1})$ & $\|\bh_2\|^2 \SNR_2 + \cO(\SNR_1^{-1})$\\
\bottomrule
\end{tabular}
\begin{tablenotes} \footnotesize
\item[*] From the symmetry between $\bh_1\sqrt{\SNR_1}$ and $\bh_2\sqrt{\SNR_2}$ in all cases, there is no need of presenting the results for $\SNR_2 \rightarrow 0^+$ and $\SNR_2\rightarrow \infty$.
\end{tablenotes}
\end{threeparttable}
\end{table*}

\subsection{Comparison with AF Relays Continued}

In this section, we resume the comparison of active RIS and traditional AF relays. Two configurations of the active RIS are considered. One is an equal-amplitude configuration
\begin{align}
    p_1 = \ldots = p_N =p, \label{equalAmplitude}
\end{align}
which is considered for the sake of analysis in \cite{zhi2022active,zhang2022active},
and the other is the optimal configuration \eqref{closedFormAmplitude}. 
The optimal amplitude for the equal-amplitude case is (see Sec. \ref{subarrayConstraint})
\begin{align}
    p = \frac{1}{\sqrt{\sum_n \gamma_n}} = \sqrt{\frac{\SNR_2}{\|\bh_1\|^2\SNR_1+N}}
\end{align}
and the corresponding receive SNR is
\begin{align}
    &\frac{(\sum_n \alpha_n)^2 p^2}{1+ (\sum_n \beta_n) p^2}\SNR_1\\
    % &=\frac{(\sum_n \alpha_n)^2}{\sum_n(\beta_n+\gamma_n)}\SNR_1\\
    &=\bigg(\frac{(\sum_{n}[\bh_2]_n [\bh_1]_n)^2}{\|\bh_1\|^2 \SNR_1+\|\bh_2\|^2\SNR_2+N}\bigg) \SNR_1 \SNR_2. \label{receivedSnrIrsEqual}
\end{align}

Putting all together---\eqref{receivedSnrMaximum}, \eqref{receivedSnrIrsMaximum}, and \eqref{receivedSnrIrsEqual}, specifically---three architectures are compared in Table \ref{comparisonOfReceivedSnr} in terms of receive SNR (see App. \ref{consistencyCheckProof} for consistency check). Optimally configured active RISs cannot perform as well as AF relays if and only if $[\bh_1]_n[\bh_2]_n \neq 0$ for at most one $n$ (refer to App. \ref{consistencyCheckProof} again), which is essentially identical to the case of single RIS element.
Interestingly, this performance gap vanishes when $\SNR_1 \rightarrow \infty$ or $\SNR_2 \rightarrow \infty$.

\begin{figure*}
    \centering
    \subfloat[$\SNR_2 = -10$ dB]
    {
        \includegraphics[width=0.32\linewidth]{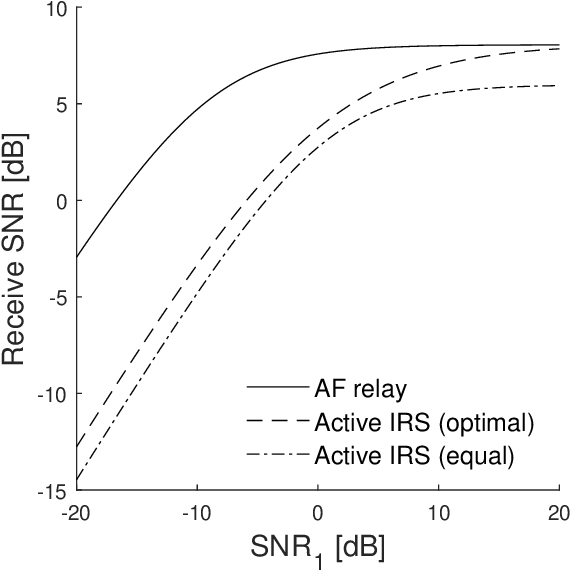}
    }
    \subfloat[$\SNR_2 = 0$ dB]
    {
        \includegraphics[width=0.32\linewidth]{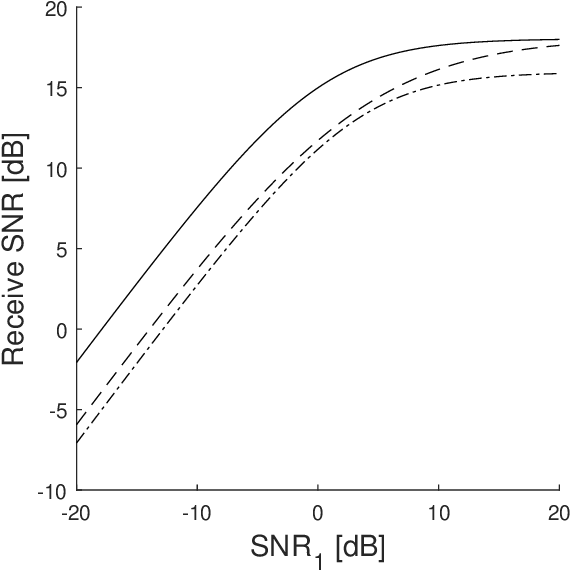}
    }
    \subfloat[$\SNR_2 = 10$ dB]
    {
        \includegraphics[width=0.32\linewidth]{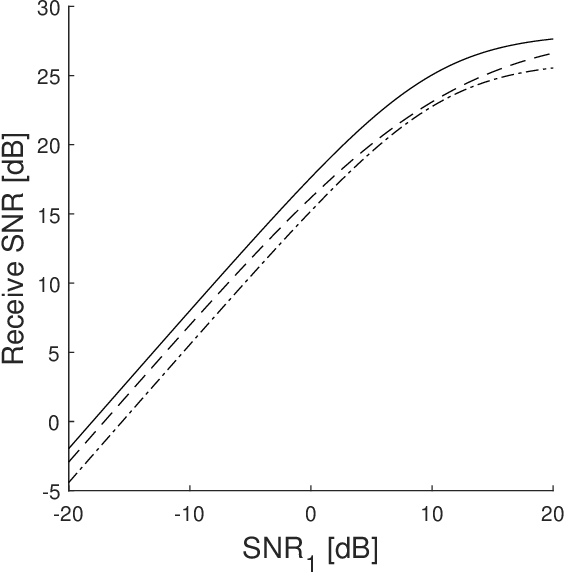}
    }
    \caption{Comparison of receive SNR averaged over thousand channel realizations with $N=64$.}
    \label{comparisonThreeArchitectures}
\end{figure*}

Fig. \ref{comparisonThreeArchitectures} compares the receive SNR with a variety of setups. Rayleigh fading channels, i.e., $\bh_1, \bh_2 \sim \cN_\bbC(0,N\bI)$, are considered. Large scale effects are absorbed into $\SNR_1$ and $\SNR_2$. We do not adopt this definition in the first place (recall \eqref{snrDefinition}) since it restricts the channel model.

\section{Optimal Amplitudes When Direct Link Exists}
\label{whenDirectLinkExists}

% We assume $h_{\rm d} \neq 0$; the case $h_{\rm d}=0$ will be addressed in Sec \ref{whenDirectLinkDoesNotExist}. Also, there is no loss of generality in assuming $[\bh_1]_n\neq 0$ and $[\bh_2]_n\neq 0$ for all $n$; if $[\bh_1]_n = 0$ or $[\bh_2]_n = 0$, setting $p_n=0$ for such $k$ never lowers the objective without violating the constraint.

In this sesction, we extend the result to $h_{\rm d}\neq 0$. Since the case $\sum_n \frac{\alpha_n^2 \gamma_n}{\beta_n^2} \leq h_{\rm d}^2$
is already solved in Sec. \ref{uniqueStationaryPoint}, we consider the remaining case $\sum_n \frac{\alpha_n^2 \gamma_n}{\beta_n^2} > h_{\rm d}^2$.

\subsection{Optimal Amplitudes}
On top of the argument in Sec. \ref{whenDirectLinkDoesNotExist}, we need an additional trick in this case. Splitting one in the denominator into $\frac{1}{1+\eta}$ and $\frac{\eta}{1+\eta}$ with some constant $\eta>0$, and repeating the argument in Sec. \ref{whenDirectLinkDoesNotExist} gives
\begin{align}
    \frac{(h_{\rm d}+\sum_n \alpha_n p_n)^2}{1+ \sum_n \beta_n p_n^2} &\leq  \frac{(h_{\rm d} + \sum_n \alpha_n p_n)^2}{\frac{1}{1+\eta} + \sum_n (\beta_n+\frac{\eta}{1+\eta}\gamma_n) p_n^2}\\
    &\leq (1+\eta)h_{\rm d}^2 + \sum_n \frac{\alpha_n^2}{\beta_n+\frac{\eta}{1+\eta}\gamma_n}, \label{cleverCauchy}
\end{align}
Since the inequality holds for any $\eta >0$, one can optimize the bound by minimizing the right-hand side with respect to $\eta$. Differentiating the right-hand side with respect to $\eta>0$ and equating it with zero, we obtain
\begin{align}
    f(\eta) \equiv \sum_n \frac{\alpha_n^2\gamma_n}{(\beta_n + \eta(\beta_n+\gamma_n))^2} - h_{\rm d}^2 = 0. \label{rootFinding}
\end{align}
Each summand is decreasing, $f(\eta) \rightarrow -h_{\rm d}^2 <0$ as $\eta\rightarrow \infty$, and
\begin{align}
    f(0) = \sum_n \frac{\alpha_n^2\gamma_n}{\beta_n^2} - h_{\rm d}^2 > 0
\end{align}
from the assumption made in the beginning of the section. Therefore, the solution $\eta^*$ of \eqref{rootFinding} uniquely exists.

To determine whether we can attain this bound or not, we need to investigate the equality condition of \eqref{cleverCauchy}, which is
\begin{align}
    \sum_n \gamma_n p_n^2 -1 = 0 \qquad p_n = \frac{1}{h_{\rm d}}\cdot  \frac{\alpha_n}{\beta_n + \eta(\beta_n+\gamma_n)},
\end{align}
respectively. Plugging the latter one into the former one, we happen to obtain \eqref{rootFinding}. It implies that we can attain the bound
\begin{align}
    (1+\eta^*)h_{\rm d}^2 + \sum_n \frac{\alpha_n^2}{\beta_n+\frac{\eta^*}{1+\eta^*}\gamma_n},
\end{align}
proving the global optimality.

\subsection{Numerical Computation}
The remaining task is to solve the root finding problem \eqref{rootFinding}.
The convexity of $f$ comes in handy from a computational standpoint.
Let us consider Newton's method, that is,
\begin{align}
    \eta^{(t+1)}\gets \eta^{(t)} - \frac{f(\eta^{(t)})}{f'(\eta^{(t)})}
\end{align}
where $t$ is the iteration count. For any initial point $\eta^{(0)}<\eta^*$, $\eta^{(t)}$ is monotonically increasing and converges quadratically to $\eta^*$ \cite[Exer. 22.14]{spivak1994calculus}.

One may use $\eta^{(0)} = 0$, however it can be problematic for very small $h_{\rm d}$. Better initial guess can be obtained from the following observation:
\begin{align}
    \sum_n \frac{\alpha_n^2\gamma_n}{(\eta+1)^2(\beta_n+\gamma_n)^2} < f(\eta) < \sum_n \frac{\alpha_n^2\gamma_n}{\eta^2(\beta_n+\gamma_n)^2}.
\end{align}
From this observation, it is clear that $\eta^*$ lies in the unit-length interval
\begin{align}
    \Bigg(\frac{1}{h_{\rm d}}\sqrt{\sum_n \frac{\alpha_n^2\gamma_n}{(\beta_n+\gamma_n)^2}}-1, \frac{1}{h_{\rm d}}\sqrt{\sum_n \frac{\alpha_n^2\gamma_n}{(\beta_n+\gamma_n)^2}}\Bigg).
\end{align}
In this respect,
\begin{align}
    \eta^{(0)} = \max\Bigg(0,\frac{1}{h_{\rm d}}\sqrt{\sum_n \frac{\alpha_n^2\gamma_n}{(\beta_n+\gamma_n)^2}}-1\Bigg)
\end{align}
is a decent initial guess for Newton's method with convergence guarantee.
Note that each Newton step, evaluation of $f(\eta^{(t)})$ and $f'(\eta^{(t)})$, requires $\cO(N)$ computations.

Incidentally, the very problem, computing the inverse of $f$, appears in trust region subproblem. Interested readers are referred to \cite{gander1989constrained}.

\subsection{Relation to the Case Where Direct Link Does Not Exist}
% That said, we can easily extend the result to the case where $h_{\rm d}=0$ without repeating the entire derivation.
We can alternatively obtain the result in Sec. \ref{whenDirectLinkDoesNotExist} by letting $h_{\rm d} \rightarrow 0^+$. In doing so, the condition
\begin{align}
    \sum_n\frac{\alpha_n^2\gamma_n}{\beta_n^2} > h_{\rm d}^2
\end{align}
is eventually satisfied. Thus, $\eta$ should satisfy \eqref{rootFinding}, which can be rewritten as
\begin{align}
    \frac{\eta^2}{\sum_n \frac{\alpha_n^2\gamma_n}{(\frac{\beta_n}{\eta} +\beta_n+\gamma_n)^2}} =  \frac{1}{h_{\rm d}^2}.
\end{align}
Intuition suggests
\begin{align}
    \eta = \bigg(\!\sum_n \frac{\alpha_n^2\gamma_n}{(\beta_n +\gamma_n)^2}\!\bigg)^{\frac{1}{2}} \frac{1}{h_{\rm d}} + o\bigg(\frac{1}{h_{\rm d}}\bigg),
\end{align}
which can be formally proved with Lagrange inversion theorem \cite[Thm. 2.3.1]{krantz2002implicit}.
This rediscovers \eqref{closedFormAmplitude}:
\begin{align}
    p_k &= \frac{1}{h_{\rm d}}\cdot  \frac{\alpha_n}{\beta_n + \Big(\big(\sum_n \frac{\alpha_n^2\gamma_n}{(\beta_n +\gamma_n)^2}\big)^{\frac{1}{2}} \frac{1}{h_{\rm d}} + o\big(\frac{1}{h_{\rm d}}\big)\Big)(\beta_n+\gamma_n)} \nonumber\\ 
    &\rightarrow \frac{\alpha_n}{\beta_k+\gamma_k} \bigg(\!\sum_n \frac{\alpha_n^2\gamma_n}{(\beta_n +\gamma_n)^2}\!\bigg)^{\!-\frac{1}{2}}.
\end{align}

% \subsection{Power at the RIS}
% \begin{align}
%     \sum_n ([\bh_1]_n^2 \SNR_1+1) p_n^2 = \min(\SNR_2,...)
% \end{align}

% Consider the case \eqref{highSnrCondition}.
% Optimal:
% \begin{align}
%     (h_{\rm d}^2+\|\bh_1\|^2)\SNR_1
% \end{align}
% Closed-form

% \subsubsection{Comparison With Amplify-And-Forward Relay}
% If the diagonal constraint is omitted, the optimal structure of $\bPhi$ is
% \begin{align}
%     \bPhi \propto \bh_2
% \end{align}

% \subsection{Infinite Power Budget at the Transmitter}
% \begin{align}
%      p_k =\frac{[\bh_2]_k }{[\bh_1]_k \sqrt{\SNR_1}}\bigg(\!\sum_n [\bh_2]_n^2\!\bigg)^{\!-\frac{1}{2}} \sqrt{\SNR_2}
% \end{align}

% \begin{align}
%     ([\bh_1]_k^2 \SNR_1+1) p_k^2  = [\bh_2]_k^2 \bigg(\!\sum_n [\bh_2]_n^2\!\bigg)^{\!-1} \SNR_2
% \end{align}

% \subsection{Infinite Power Budget at the RIS}
% \begin{align}
%     p_k =\frac{[\bh_1]_k}{[\bh_2]_k}\bigg(\!\sum_n \frac{[\bh_1]_n^2}{[\bh_2]_n^2}([\bh_1]_n^2\SNR_1+1)\!\bigg)^{\!-\frac{1}{2}}\sqrt{\SNR_2} + o(\sqrt{\SNR_2})
% \end{align}

% \begin{align}
%     ([\bh_1]_k^2 \SNR_1+1) p_k^2 =\frac{[\bh_1]_k^2}{[\bh_2]_k^2}\bigg(\!\sum_n \frac{[\bh_1]_n^2}{[\bh_2]_n^2}\!\bigg)^{-1}\SNR_2 + o(\SNR_2)
% \end{align}
% % \section{Asymptotic Behavior for Rayleigh Fading}

% \heedong{ADD A CONCRETE EXAMPLE WITH FIGURE!}

\begin{table}[]
\centering
\begin{threeparttable}
\caption{Assessment of Prior-Art Algorithms}
\label{algorithmComparison}
\begin{tabular}{cccccc} 
\toprule
\multicolumn{3}{c}{Parameters} & \multicolumn{3}{c}{Average receive SNR [dB]} \\
$\SNR_1$ [dB] & $\SNR_2$ [dB] & $h_{\rm d}$ & Optimum & SCA & AO\\
\midrule
$-$10 & $-$10 & 0 & $-$15.6431 & $-$15.6431 & $-$15.6431 \\
$-$10 & $-$10 & 1 & $-$6.7920 & $-$6.7920 & $-$6.8048\\
$-$10 &  10 & 0 & $-$5.1313 & $-$5.1313 & $-$5.1313 \\
$-$10 &  10 & 1 & $-$3.1491 & $-$3.1491 & $-$3.5859 \\
 10 & $-$10 & 0 & $-$5.2221 & $-$5.2221 & $-$5.2221 \\
 10 & $-$10 & 1 & 11.2521 & 11.2521 & 11.2477 \\
 10 &  10 & 0 & 10.8216 & 10.8216 & 10.8216 \\
 10 &  10 & 1 & 15.3245 & 15.3245 & 15.0697 \\
\bottomrule
\end{tabular}
\end{threeparttable}
\end{table}

\section{Assessing Existing Algorithms}
\label{assessingExistingAlgorithms}
Having known the global optimum, we can assess the existing algorithms, sequential convex approximation (SCA) \cite{long2021active} and alternating optimization (AO) \cite{zhang2022active}, through numerical simulations. Initial point of both algorithms is chosen as $p_n = 1/\sqrt{\sum_n \gamma_n}$.

As in Sec. \ref{whenDirectLinkDoesNotExist}, Rayleigh fading channels are considered, that is, $\bh_1, \bh_2 \sim \cN_\bbC(0,N\bI)$, and large scale effects are absorbed into $\SNR_1$ and $\SNR_2$. Table \ref{algorithmComparison} presents the simulation results with $N=4$. The receive SNR is averaged over a hundred of channel realizations. The numerical results strongly suggests that, at least in these settings, both prior-art algorithms somehow attain the global optimum when the direct link is blocked. Interestingly, when the direct link exists, AO exhibits performance degradation while SCA does not.

\section{Incorporating Additional RIS Constraints}
\label{incormporatingAdditionalRisConstraints}
The system model in Sec. \ref{systemModel} is a mathematical description of the active RIS in the real world. The tractability of the model comes at the cost of loss in reality. Towards more realistic model, similar to the practical considerations in traditional antenna arrays \cite{park2017dynamic, yu2007transmitter, vu2011miso}, some prior art in active RIS takes analogous hardware constraints into consideration.  
Two hardware constraints which do not harm the tractability will be introduced.

\subsection{Subarray Constraint}
\label{subarrayConstraint}
The system model in Sec. \ref{systemModel} presumes the use of RIS elements each having its dedicated power amplifier. To reduce the hardware complexity, an RIS architecture comprising of multiple subarrays with one amplifier per array \cite{tasci2022new} or per subarray \cite{liu2021active} is considered. 
In this case, each subarray acts like a single element and the problem boils down to the case which has been solved.

% Let $M$ be the number of subarrays.
Let us denote $\bar{p}_m$ be an amplifying factor of the $m$th subarray and $I_m$ be the set of RIS element indices connected to the $m$th subarray. With this additional constraint, the optimization problem \eqref{simplifiedProblem} can be recast as
\begin{align}
    \max_{\{\bar{p}_m\}} \quad&  \frac{(h_{\rm d} + \sum_m \bar{\alpha}_m \bar{p}_m)^2}{1+ \sum_m \bar{\beta}_m \bar{p}_m^2}
    % 2\log\bigg(1+\sum_n \alpha_n p_n\bigg)-\log\bigg(1+ \sum_n \beta_n p_n^2\bigg) 
    \\
    \text{s.t.} \quad& \sum_m  \bar{\gamma}_m\bar{p}_m^2 -1\leq 0, \nonumber
\end{align}
where
\begin{align}
    \bar{\alpha}_m = \sum_{n\in I_m} \alpha_n \qquad
    \bar{\beta}_m = \sum_{n\in I_m} \beta_n \qquad
    \bar{\gamma}_m = \sum_{n\in I_m}\gamma_n.
\end{align}
It shares the structure with the original problem, hence the algorithms in Secs. \ref{whenDirectLinkDoesNotExist} and \ref{whenDirectLinkExists} can be used without any modification.

For the sake of completeness, let us consider the case $h_{\rm d}=0$ where the closed-form solution is available. In this case, the optimal amplitudes are
\begin{align}
    p_m = \frac{\bar{\alpha}_m}{\bar{\beta}_m+\bar{\gamma}_m}\bigg(\sum_\ell \frac{\bar{\alpha}_\ell^2\bar{\gamma}_\ell}{(\bar{\beta}_\ell+\bar{\gamma}_\ell)^2}\bigg)^{-\frac{1}{2}},
\end{align}
and the corresponding receive SNR is
\begin{align}
    &\bigg(\sum_m \frac{(\sum_{n\in I_m}[\bh_2]_n [\bh_1]_n)^2}{(\sum_{n\in I_m}[\bh_1]_n^2) \SNR_1+(\sum_{n\in I_m}[\bh_2]_n^2)\SNR_2+|I_m|}\bigg)\nonumber\\
    &\qquad\qquad\qquad\qquad\qquad\qquad\qquad\qquad \cdot \SNR_1 \SNR_2,
\end{align}
where $|\cdot|$ denotes the cardinality of the set.
% With the simplest architecture with a single subarray, the receive SNR is
% \begin{align}
%     \bigg(\frac{(\sum_{n}[\bh_2]_n [\bh_1]_n)^2}{\|\bh_1\|^2 \SNR_1+\|\bh_2\|^2\SNR_2+N}\bigg) \SNR_1 \SNR_2,
% \end{align}

\subsection{Per-Element Amplitude Constraint}
In \cite{long2021active}, the per-element amplitude constraint at the RIS is additionally taken into account on top of the total power constraint. In this subsection, we only consider the per-element amplitude constraint:
\begin{align}
    \max_{p_n} \quad& \frac{(h_{\rm d}+\sum_n \alpha_n p_n)^2}{1+ \sum_n \beta_n p_n^2} \label{optimizationProblemPerElement}\\
    \text{s.t.} \quad& 0 \leq p_n \leq \bar{p}_n\quad \forall n. \nonumber
\end{align}
It is shown in App. \ref{perElementAmplitudeProof} that the optimal amplitudes are
\begin{align}
    p_k = \min\!\bigg(\frac{\alpha_k}{\beta_k}\rho,\bar{p}_k\bigg).
    % =\frac{\alpha_k}{\beta_k}\min\!\bigg(\rho, \frac{\beta_k}{\alpha_k}\bar{p}_k\bigg).
\end{align}
where $\rho$ is the unique solution of
\begin{align}
    h_{\rm d}\rho + \sum_n \bar{p}_n \max\!\big(0,\alpha_n\rho-\beta_n\bar{p}_n\big)= 1. \label{rootFindingFirstCase}
\end{align}
One can explicitly compute the root of \eqref{rootFindingFirstCase} by evaluating the left-hand side at the breakpoints. Comparing these values with one, the right-hand side, and exploiting the piecewise linear property of the left-hand side complete the computation. 

One can easily check that it is consistent with the result in Sec \ref{uniqueStationaryPoint}.

\begin{table*}[]
\centering
\begin{threeparttable}
\setlength{\tabcolsep}{10pt}
\caption{Summary of the Results}
\label{summaryTable}
\begin{tabular}{ccc} 
\toprule
& $\sum_n \frac{\alpha_n^2 \gamma_n}{\beta_n^2} \leq h_{\rm d}^2$ & $\sum_n \frac{\alpha_n^2 \gamma_n}{\beta_n^2} > h_{\rm d}^2$ \\
\midrule
$h_{\rm d} = 0$ & - & $p_k = \frac{\alpha_k}{\beta_k+\gamma_k}\big(\sum_n \frac{\alpha_n^2\gamma_n}{(\beta_n+\gamma_n)^2}\big)^{-\frac{1}{2}}$ (Sec. \ref{whenDirectLinkDoesNotExist})\\
$h_{\rm d} > 0$ & $p_k = \frac{\alpha_k}{h_{\rm d} \beta_k}$ (Sec. \ref{uniqueStationaryPoint}) & $p_k = \frac{1}{h_{\rm d}}\cdot  \frac{\alpha_k}{\beta_k + \eta(\beta_k+\gamma_k)}$ with $\sum_n \frac{\alpha_n^2\gamma_n}{(\beta_n + \eta(\beta_n+\gamma_n))^2} = h_{\rm d}^2$ (Sec. \ref{whenDirectLinkExists}) \\
\bottomrule
\end{tabular}
\end{threeparttable}
\end{table*}

\section{Conclusion}
\label{summary}

This paper has presented a numerical method of computing globally optimal configuration of an active RIS, which is summarized in Table \ref{summaryTable}. The complexity of the method is linear in the number of RIS elements which is orders-of-magnitude improvement comparing to the existing algorithms which are not even guaranteed to be globally optimal.

Throughout the paper, the availability of perfect CSI is assumed, which cannot be obtained through conventional channel estimation methods owing to the cost-efficient nature of RIS.
In this respect, one apparent worthwhile direction to explore is incorporation of channel estimation (see \cite{swindlehurst2022channel} and \cite{ren2022configuring} for explicit and implicit channel estimation methods).  

% Much work has been done in this direction with an additional sensor at the RIS or with a feedback at the receiver (see \cite{swindlehurst2022channel} for a comprehensive set of such approaches). More practically viable approach is blind beamforming \cite{ren2022configuring} which does not require explicit channel estimation just as beam training at mmWave frequencies.

% power sensing \cite{zelaya2021lava}

\appendices

\section{}
\label{relayProof}
Using Cauchy-Schwarz inequality, we have
\begin{align}
    \frac{|\bh_2^\top\bPhi\bh_1|^2}{1+\|\bh_2^\top \bPhi\|^2}\SNR_1 &\leq \frac{|\bh_2^\top\bPhi\bh_1|^2}{1+\frac{|\bh_2^\top\bPhi\bh_1|^2}{\|\bh_1\|^2}}\SNR_1\\
    &= \frac{1}{\frac{1}{\|\bh_1\|^2}+\frac{1}{|\bh_2^\top\bPhi\bh_1|^2}}\SNR_1
\end{align}
and
\begin{align}
    \SNR_2 & \geq \SNR_1\|\bPhi \bh_1\|^2  + \|\bPhi\|_{\rm F}^2 \\
    & \geq  
    \frac{\SNR_1 |\bh_2^\top \bPhi \bh_1|^2}{\|\bh_2\|^2}  + \frac{ |\bh_2^\top \bPhi \bh_1|^2}{\|\bh_1\|^2\|\bh_2\|^2}\\
    & = \frac{\|\bh_1\|^2\SNR_1 + 1}{\|\bh_1\|^2\|\bh_2\|^2}|\bh_2^\top \bPhi \bh_1|^2.
\end{align}
Combining these inequalities, we immediately have
\begin{align}
    &\frac{|\bh_2^\top\bPhi\bh_1|^2}{1+\|\bh_2^\top \bPhi\|^2}\SNR_1\nonumber\\
    &\qquad \leq \frac{1}{\frac{1}{\|\bh_1\|^2}+\frac{\|\bh_1\|^2\SNR_1 + 1}{\|\bh_1\|^2\|\bh_2\|^2\SNR_2}}\SNR_1\\
    &\qquad =\frac{\|\bh_1\|^2\|\bh_2\|^2}{\|\bh_1\|^2\SNR_1 + \|\bh_2\|^2\SNR_2+1} \SNR_1\SNR_2.
\end{align}
One can easily check that all the equalities can be simultaneously held, which concludes the proof.

\section{}
\label{fractionalProof}
Using Dinkelbach's transform \cite{dinkelbach1967nonlinear}, the problem \eqref{fractionalProgramming} can be reformulated as a series of convex optimization problems:
\begin{align}
    & (p_1^{(t+1)},\ldots,p_N^{(t+1)})\nonumber\\
    & = \argmax_{\substack{(p_1,\ldots,p_N):\\ \sum_n \gamma_n p_n^2 -1\leq 0}}\sum_n \alpha_n p_n - \lambda^{(t)} \sqrt{1+\sum_n \beta_n p_n^2} \label{fractionalStep}
 \end{align}
where
\begin{align}
    \lambda^{(t)} = \frac{1+\sum_n \alpha_n p_n^{(t)}}{\sqrt{1+\sum_n \beta_n \big(p_n^{(t)}\big)^2}}.
\end{align}
is an auxiliary variable for iteration index $t$.

Introducing an auxiliary variable, one can turn the problem into SOCP,
\begin{align}
    \max_{\{p_n\},y} \ & \begin{bmatrix}\alpha_1 & \ldots & \alpha_N & -\lambda^{(t)} \end{bmatrix}  \bp  \\
    \text{s.t.} \ & \| \diag(\sqrt{\gamma_1},\ldots,\sqrt{\gamma_N},0)\bp\| \leq 1\nonumber\\
    & \| \diag(\sqrt{\beta_1},\ldots,\sqrt{\beta_N},0)\bp+1\| \leq \begin{bmatrix}0 & \ldots & 0 & 1 \end{bmatrix} \bp \nonumber,
\end{align}
where $\bp = \begin{bmatrix}p_1 & \ldots & p_N & y \end{bmatrix}^\top$.

\section{}
\label{consistencyCheckProof}
As a byproduct of the receive SNR results, we have
\begin{align}
    &\frac{(\sum_n [\bh_1]_n [\bh_2]_n)^2}{\|\bh_1\|^2 \SNR_1+\|\bh_2\|^2\SNR_2+N}\nonumber\\
    &\qquad \leq \sum_n \frac{[\bh_1]_n^2 [\bh_2]_n^2}{[\bh_1]_n^2 \SNR_1+[\bh_2]_n^2\SNR_2+1}\nonumber\\
    &\qquad \leq \frac{\|\bh_1\|^2\|\bh_2\|^2}{\|\bh_1\|^2\SNR_1 + \|\bh_2\|^2\SNR_2+1}.
\end{align}
This section checks the consistency by proving these inequalities in alternative ways.

The first inequality can be alternatively derived from Cauchy-Schwarz inequality:
\begin{align}
    &(\|\bh_1\|^2 \SNR_1+\|\bh_2\|^2\SNR_2+N)\nonumber\\
    &\qquad\qquad \cdot \bigg(\sum_n \frac{[\bh_1]_n^2 [\bh_2]_n^2}{[\bh_1]_n^2 \SNR_1+[\bh_2]_n^2\SNR_2+1}\bigg)\\
    &=\bigg(\sum_n [\bh_1]_n^2 \SNR_1+[\bh_2]_n^2\SNR_2+1 \bigg)\nonumber\\
    &\qquad\qquad \cdot \bigg(\sum_n \frac{[\bh_1]_n^2 [\bh_2]_n^2}{[\bh_1]_n^2 \SNR_1+[\bh_2]_n^2\SNR_2+1}\bigg)\\
    &\leq \bigg(\sum_n [\bh_1]_n [\bh_2]_n\bigg)^2.
\end{align}

It is the second inequality which is more obscure, which essentially says the function
\begin{align}
    (x,y) \mapsto \frac{xy}{1+x+y}
\end{align}
defined over $[0,\infty)\times [0,\infty)$ is superadditive.
It is sufficient to show
\begin{align}
    \frac{x_1y_1}{1+x_1+y_1} \!+\! \frac{x_2y_2}{1+x_2+y_2} \!\leq\! \frac{(x_1+x_2)(y_1+y_2)}{1+x_1+y_1+x_2+y_2},
\end{align}
which is equivalent to
\begin{align}
    &(x_1+x_2)(y_1+y_2)(1+x_1+y_1)(1+x_2+y_2) \nonumber \\
    &\quad - x_1y_1(1+x_2+y_2)(1+x_1+y_1+x_2+y_2) \nonumber \\
    &\quad - x_2y_2(1+x_1+y_1)(1+x_1+y_1+x_2+y_2) \geq 0. \label{superaddivity1}
\end{align}
The left-hand side of \eqref{superaddivity1} can be simplified to
\begin{align}
    (x_1y_2-x_2y_1)^2 + x_1y_2(x_1+y_2+1) + x_2 y_1(x_2+y_1+1),
\end{align}
which is clearly nonnegative. The equality condition is therefore $x_1y_2 = x_2y_1 = 0$. Using this, one can easily deduce that the equality condition of 
\begin{align}
    \sum_n\frac{x_ny_n}{1+x_n+y_n}\leq \frac{(\sum_n x_n)(\sum_n y_n)}{1+\sum x_n + \sum_n y_n},
\end{align}
is when the number of positive elements in $\{x_ny_n\}$ is at most one.

\section{}
\label{perElementAmplitudeProof}
To solve \eqref{optimizationProblemPerElement} explicitly, we first apply the log function to the objective and write down the Karush-Kuhn-Tucker (KKT) conditions:
\begin{align}
    &\frac{\alpha_k}{h_{\rm d}+\sum_n \alpha_n p_n}-\frac{\beta_k p_k}{1+\sum_n \beta_n p_n^2}- \mu_k = 0 \label{stationarity}\\
    & 0\leq p_k \leq \bar{p}_k, \label{primalFeasibility}\\
    & \mu_k \geq 0 \label{dualFeasibility}\\
    & \mu_k(p_k-\bar{p}_k) = 0. \label{slackness}
\end{align}
Eliminating $\mu_k$, \eqref{stationarity} and \eqref{slackness} become
\begin{align}
    &\frac{\alpha_k}{h_{\rm d}+\sum_n \alpha_n p_n}-\frac{\beta_k p_k}{1+\sum_n \beta_n p_n^2}\geq 0, \label{stationarySimplified}\\
    & \bigg(\!\frac{\alpha_k}{h_{\rm d}\!+\!\sum_n \alpha_n p_n}-\frac{\beta_k p_k}{1\!+\!\sum_n \beta_n p_n^2}\!\bigg)(p_k-\bar{p}_k) = 0, \label{slacknessSimplified}
\end{align}
respectively. 
Denoting
\begin{align}
    \rho = \frac{1+\sum_n \beta_n p_n^2}{h_{\rm d} + \sum_n \alpha_n p_n}, \label{rho}
\end{align}
we can rewrite \eqref{stationarySimplified} and \eqref{slacknessSimplified} as
\begin{align}
    p_k \leq \frac{\alpha_k}{\beta_k} \rho \qquad \bigg(\frac{\alpha_k}{\beta_k}\rho-p_k\bigg)(p_k-\bar{p}_k) = 0,
\end{align}
respectively. It is equivalent to
\begin{align}
    p_k = \min\!\bigg(\frac{\alpha_k}{\beta_k}\rho,\bar{p}_k\bigg).
    % =\frac{\alpha_k}{\beta_k}\min\!\bigg(\rho, \frac{\beta_k}{\alpha_k}\bar{p}_k\bigg).
\end{align}
Plugging it into \eqref{rho}, we obtain
\begin{align}
    \rho = \frac{1+\sum_n \beta_n \big(\min\!\big(\frac{\alpha_n}{\beta_n}\rho,\bar{p}_n\big)\big)^2}{h_{\rm d}+\sum_n \alpha_n \min\!\big(\frac{\alpha_n}{\beta_n}\rho,\bar{p}_n\big)}
\end{align}
Rearranging terms and simplifying it, we obtain
\begin{align}
    h_{\rm d}\rho + \sum_n \bar{p}_n \max\!\big(0,\alpha_n\rho-\beta_n\bar{p}_n\big)= 1.
\end{align}

\section{}
\label{alternatingOptimizationDescription}

This section describes the AO algorithm proposed in \cite{zhang2022active}.
Although this algorithm is designed for more general multi-user MISO setup, it can be applied to SISO case. It is worth rewriting the algorithm in great detail as it is unclear that which part of the algorithm could be pruned when narrowing down the scope to SISO.

Let us first rewrite the problem as
\begin{align}
    \max_{\{p_n\}} \quad& \log \bigg(\!1+\frac{(h_{\rm d}+\sum_n \alpha_n p_n)^2}{1+ \sum_n \beta_n p_n^2}\!\bigg) \nonumber\\
    \text{s.t.} \quad& \sum_n \gamma_n p_n^2 -1\leq 0.  
\end{align}
From the concavity of the log function, we have \eqref{tangentLine} for any $\rho\in\bbR$ as concave function always lie below its tangent lines. Note that slightly modifying this gives an alternative constructive derivation of \cite[Thm. 3]{shen2018fractional2}.
\begin{figure*}
\begin{align}
    \log \bigg(1+\frac{(h_{\rm d}+\sum_n \alpha_n p_n)^2}{1+ \sum_n \beta_n p_n^2}\bigg)  + \frac{(1+ \sum_n \beta_n p_n^2)}{(h_{\rm d}+\sum_n \alpha_n p_n)^2+(1+ \sum_n \beta_n p_n^2)} \bigg(\rho - \frac{(h_{\rm d}+\sum_n \alpha_n p_n)^2}{1+ \sum_n \beta_n p_n^2}\bigg) \geq \log(1+\rho) \label{tangentLine}
\end{align}
\hrulefill
\end{figure*}
The problem of interest is now
\begin{align}
    \max_{\{p_n\},\rho} \ & \log(1+\rho) - \rho  + \frac{(1+\rho)(h_{\rm d}+\sum_n \alpha_n p_n)^2}{(h_{\rm d}+\sum_n \alpha_n p_n)^2+ (1+\sum_n \beta_n p_n^2)} \nonumber\\
    \text{s.t.} \ & \sum_n \gamma_n p_n^2 -1\leq 0 
\end{align}
with an introduction of the auxiliary variable $\rho\in\bbR$.
\begin{figure*}
\vspace{-4mm}
\begin{align}
    \bigg(\!\bigg(\!h_{\rm d}\!+\!\sum_n \alpha_n p_n\!\bigg)^{\!2}\!+\! \bigg(\!1+\sum_n \beta_n p_n^2\!\bigg)\!\bigg)w^2\!-\!2\sqrt{1+\rho}\bigg(\!h_{\rm d}\!+\!\sum_n \alpha_n p_n\!\bigg) w \!+\! \frac{(1+\rho)(h_{\rm d}+\sum_n \alpha_n p_n)^2}{(h_{\rm d}+\sum_n \alpha_n p_n)^2+ (1+\sum_n \beta_n p_n^2)}\geq 0 \label{yetAnotherAuxiliaryVariable}
\end{align}
\hrulefill
\end{figure*}

We proceed by repeating a similar argument.
From \eqref{yetAnotherAuxiliaryVariable} which holds for any $w\in\bbR$, the problem could be transformed into
\begin{align}
    \max_{\{p_n\},\rho,w} \ & \log(1+\rho) -\rho + 2\sqrt{1+\rho}\bigg(\!h_{\rm d}+\sum_n \alpha_n p_n\!\bigg) w\nonumber\\
    &\qquad - \bigg(\!\bigg(\!h_{\rm d}+\sum_n \alpha_n p_n\!\bigg)^{\!2}\!+\! \bigg(\!1+\sum_n \beta_n p_n^2\!\bigg)\!\bigg)w^2 \nonumber\\
    \text{s.t.} \ & \sum_n \gamma_n p_n^2 -1\leq 0.
\end{align}

The transformed problem is tackled with alternating optimization iterating three procedures: fixing two of $\{p_n\}$, $\rho$, and $w$, and optimizing the other. Fixing $\{p_n\}$ and $w$, simple calculus gives the optimal $\rho$:
\begin{align}
    \rho \gets \frac{\xi^2+\xi\sqrt{\xi^2+4}}{2}
\end{align}
with $\xi = (h_{\rm d}+\sum_n \alpha_n p_n)w$. Similarly, fixing $\{p_n\}$ and $\rho$, it is a quadratic function of $w$ and the optimal $w$ equals
\begin{align}
    w \gets \frac{\sqrt{1+\rho}(h_{\rm d}+\sum_n \alpha_n p_n)}
    {(h_{\rm d}+\sum_n \alpha_n p_n)^{2}+(1+\sum_n \beta_n p_n^2)}
\end{align}
Lastly, with $\rho$ and $w$ being fixed, omitting irrelevant constants, the problem is essentially
\begin{align}
    \max_{\{p_n\}} \ & 2\sqrt{1+\rho}\bigg(\!h_{\rm d}+\sum_n \alpha_n p_n\!\bigg) w\nonumber\\
    &\qquad - \bigg(\!\bigg(\!h_{\rm d}+\sum_n \alpha_n p_n\!\bigg)^{\!2}\!+\! \bigg(\!1+\sum_n \beta_n p_n^2\!\bigg)\!\bigg)w^2  \nonumber\\
    \text{s.t.} \ & \sum_n \gamma_n p_n^2 -1\leq 0, \label{trustRegion}
\end{align}
which falls into a class of trust region subproblem.

\bibliographystyle{IEEEtran}
\bibliography{jour_short,conf_short,ref}

\end{document}

%% file: input.tex
\def\bb0{{\mathbb{0}}}

\def\bb{{\boldsymbol{b}}}

\def\bh{{\boldsymbol{h}}}

\def\bp{{\boldsymbol{p}}}

\def\bv{{\boldsymbol{v}}}

\def\by{{\boldsymbol{y}}}

\def\b0{{\boldsymbol{0}}}

\def\bI{{\boldsymbol{I}}}

\def\b{{\mathrm{b}}}

\def\r0{{\mathbf{0}}}

\def\bbC{{\mathbb{C}}}

\def\bbE{{\mathbb{E}}}

\def\bbR{{\mathbb{R}}}

\def\cN{\mathcal{N}}
\def\cO{\mathcal{O}}

\def\bPhi{\bm \Phi}
\def\bsf0{{\bm{\mathsf{0}}}}

\def\N0{{N_{\mathrm{0}}}}

\def\bsf{{\boldsymbol{s}_\mathrm{f}}}

\newcommand{\be}{\begin{equation}}
\newcommand{\ee}{\end{equation}}
\newcommand{\bal}{\begin{align}}
\newcommand{\eal}{\end{align}}

\def\SNR    {{\mathsf{SNR}}}

\def\diag   {\mbox{\rm diag}}

%% file: double_0th.bbl
\begin{thebibliography}{10}
\providecommand{\url}[1]{#1}
\csname url@samestyle\endcsname
\providecommand{\newblock}{\relax}
\providecommand{\bibinfo}[2]{#2}
\providecommand{\BIBentrySTDinterwordspacing}{\spaceskip=0pt\relax}
\providecommand{\BIBentryALTinterwordstretchfactor}{4}
\providecommand{\BIBentryALTinterwordspacing}{\spaceskip=\fontdimen2\font plus
\BIBentryALTinterwordstretchfactor\fontdimen3\font minus
  \fontdimen4\font\relax}
\providecommand{\BIBforeignlanguage}[2]{{%
\expandafter\ifx\csname l@#1\endcsname\relax
\typeout{** WARNING: IEEEtran.bst: No hyphenation pattern has been}%
\typeout{** loaded for the language `#1'. Using the pattern for}%
\typeout{** the default language instead.}%
\else
\language=\csname l@#1\endcsname
\fi
#2}}
\providecommand{\BIBdecl}{\relax}
\BIBdecl

\bibitem{liaskos2018new}
C.~Liaskos, S.~Nie, A.~Tsioliaridou, A.~Pitsillides, S.~Ioannidis, and
  I.~Akyildiz, ``A new wireless communication paradigm through
  software-controlled metasurfaces,'' \emph{IEEE Commun. Mag.}, vol.~56, no.~9,
  pp. 162--169, 2018.

\bibitem{wu2019towards}
Q.~Wu and R.~Zhang, ``Towards smart and reconfigurable environment: Intelligent
  reflecting surface aided wireless network,'' \emph{IEEE Commun. Mag.},
  vol.~58, no.~1, pp. 106--112, 2019.

\bibitem{basar2019wireless}
E.~Basar, M.~Di~Renzo, J.~De~Rosny, M.~Debbah, M.-S. Alouini, and R.~Zhang,
  ``Wireless communications through reconfigurable intelligent surfaces,''
  \emph{IEEE Access}, vol.~7, pp. 116\,753--116\,773, 2019.

\bibitem{do2022line}
H.~Do, N.~Lee, and A.~Lozano, ``Line-of-sight {MIMO} via intelligent reflecting
  surface,'' \emph{IEEE Trans. Wireless Commun.}, 2022.

\bibitem{dunna2020scattermimo}
M.~Dunna, C.~Zhang, D.~Sievenpiper, and D.~Bharadia, ``{ScatterMIMO}: Enabling
  virtual {MIMO} with smart surfaces,'' in \emph{Proc. Annu. Int. Conf. Mobile
  Comput. Netw.}, 2020, pp. 1--14.

\bibitem{wu2021intelligent}
Q.~Wu, S.~Zhang, B.~Zheng, C.~You, and R.~Zhang, ``Intelligent reflecting
  surface-aided wireless communications: A tutorial,'' \emph{IEEE Trans.
  Commun.}, vol.~69, no.~5, pp. 3313--3351, 2021.

\bibitem{abari2017enabling}
O.~Abari, D.~Bharadia, A.~Duffield, and D.~Katabi, ``Enabling high-quality
  untethered virtual reality,'' in \emph{USENIX Symp. Networked Syst. Des.
  Implementation}, 2017, pp. 531--544.

\bibitem{you2021wireless}
C.~You and R.~Zhang, ``Wireless communication aided by intelligent reflecting
  surface: Active or passive?'' \emph{IEEE Wireless Commun. Lett.}, vol.~10,
  no.~12, pp. 2659--2663, 2021.

\bibitem{long2021active}
R.~Long, Y.-C. Liang, Y.~Pei, and E.~G. Larsson, ``Active reconfigurable
  intelligent surface-aided wireless communications,'' \emph{IEEE Trans.
  Wireless Commun.}, vol.~20, no.~8, pp. 4962--4975, 2021.

\bibitem{zhi2022active}
K.~Zhi, C.~Pan, H.~Ren, K.~K. Chai, and M.~Elkashlan, ``Active {RIS} versus
  passive {RIS}: Which is superior with the same power budget?'' \emph{IEEE
  Commun. Lett.}, vol.~26, no.~5, pp. 1150--1154, 2022.

\bibitem{zhang2022active}
Z.~Zhang, L.~Dai, X.~Chen, C.~Liu, F.~Yang, R.~Schober, and H.~V. Poor,
  ``Active {RIS} vs. passive {RIS}: Which will prevail in {6G}?'' \emph{IEEE
  Trans. Commun.}, 2022.

\bibitem{shu2023three}
F.~Shu, J.~Liu, Y.~Lin, Y.~Liu, Z.~Chen, X.~Wang, R.~Dong, and J.~Wang, ``Three
  high-rate beamforming methods for active {IRS}-aided wireless network,''
  \emph{IEEE Trans. Veh. Technol.}, 2023.

\bibitem{lin2023enhanced}
Y.~Lin, F.~Shu, R.~Dong, R.~Chen, S.~Feng, W.~Shi, J.~Liu, and J.~Wang,
  ``Enhanced-rate iterative beamformers for active {IRS}-assisted wireless
  communications,'' \emph{IEEE Wireless Commun. Lett.}, 2023.

\bibitem{rao2023active}
J.~Rao, Y.~Zhang, S.~Tang, Z.~Li, C.-Y. Chiu, and R.~Murch, ``An active
  reconfigurable intelligent surface utilizing phase-reconfigurable reflection
  amplifiers,'' \emph{IEEE Trans. Microw. Theory Techn.}, 2023.

\bibitem{bjornson2019intelligent}
E.~Bj{\"o}rnson, {\"O}.~{\"O}zdogan, and E.~G. Larsson, ``Intelligent
  reflecting surface versus decode-and-forward: How large surfaces are needed
  to beat relaying?'' \emph{IEEE Wireless Commun. Lett.}, vol.~9, no.~2, pp.
  244--248, 2019.

\bibitem{yu2019miso}
X.~Yu, D.~Xu, and R.~Schober, ``{MISO} wireless communication systems via
  intelligent reflecting surfaces,'' in \emph{IEEE/CIC Int. Conf. Commun.
  China}.\hskip 1em plus 0.5em minus 0.4em\relax IEEE, 2019, pp. 735--740.

\bibitem{zhang2020capacity}
S.~Zhang and R.~Zhang, ``Capacity characterization for intelligent reflecting
  surface aided {MIMO} communication,'' \emph{IEEE J. Sel. Areas Commun.},
  vol.~38, no.~8, pp. 1823--1838, 2020.

\bibitem{perovic2021achievable}
N.~S. Perovi{\'c}, L.-N. Tran, M.~Di~Renzo, and M.~F. Flanagan, ``Achievable
  rate optimization for {MIMO} systems with reconfigurable intelligent
  surfaces,'' \emph{IEEE Trans. Wireless Commun.}, 2021.

\bibitem{wu2019intelligent}
Q.~Wu and R.~Zhang, ``Intelligent reflecting surface enhanced wireless network
  via joint active and passive beamforming,'' \emph{IEEE Trans. Wireless
  Commun.}, vol.~18, no.~11, pp. 5394--5409, 2019.

\bibitem{guo2020weighted}
H.~Guo, Y.-C. Liang, J.~Chen, and E.~G. Larsson, ``Weighted sum-rate
  maximization for reconfigurable intelligent surface aided wireless
  networks,'' \emph{IEEE Trans. Wireless Commun.}, vol.~19, no.~5, pp.
  3064--3076, 2020.

\bibitem{pan2020multicell}
C.~Pan, H.~Ren, K.~Wang, W.~Xu, M.~Elkashlan, A.~Nallanathan, and L.~Hanzo,
  ``Multicell {MIMO} communications relying on intelligent reflecting
  surfaces,'' \emph{IEEE Trans. Wireless Commun.}, vol.~19, no.~8, pp.
  5218--5233, 2020.

\bibitem{xie2020max}
H.~Xie, J.~Xu, and Y.-F. Liu, ``Max-min fairness in {IRS}-aided multi-cell
  {MISO} systems with joint transmit and reflective beamforming,'' \emph{IEEE
  Trans. Wireless Commun.}, vol.~20, no.~2, pp. 1379--1393, 2020.

\bibitem{liu2021two}
X.~Liu, C.~Sun, and E.~A. Jorswieck, ``Two-user {SINR} region for
  reconfigurable intelligent surface aided downlink channel,'' in \emph{IEEE
  Int. Conf. Commun. Workshops}.\hskip 1em plus 0.5em minus 0.4em\relax IEEE,
  2021, pp. 1--6.

\bibitem{tang2007optimal}
X.~Tang and Y.~Hua, ``Optimal design of non-regenerative {MIMO} wireless
  relays,'' \emph{IEEE Trans. Wireless Commun.}, vol.~6, no.~4, pp. 1398--1407,
  2007.

\bibitem{fang2006joint}
Z.~Fang, Y.~Hua, and J.~C. Koshy, ``Joint source and relay optimization for a
  non-regenerative {MIMO} relay,'' in \emph{IEEE Workshop Sensor Array
  Multichannel Process.}, 2006, pp. 239--243.

\bibitem{taricco2022information}
G.~Taricco, ``Information rate optimization for the non-regenerative linear
  {MIMO} relay channel with a direct link and variable duty cycle,'' \emph{IEEE
  Trans. Inform. Theory}, vol.~68, no.~9, pp. 5889--5900, 2022.

\bibitem{rosenholtz1985only}
I.~Rosenholtz and L.~Smylie, ``“{The} only critical point in town” test,''
  \emph{Math. Mag.}, vol.~58, no.~3, pp. 149--150, 1985.

\bibitem{bharadia2014fastforward}
D.~Bharadia and S.~Katti, ``Fastforward: Fast and constructive full duplex
  relays,'' \emph{ACM SIGCOMM Comput. Commun. Rev.}, vol.~44, no.~4, pp.
  199--210, 2014.

\bibitem{zelaya2021lava}
R.~I. Zelaya, W.~Sussman, J.~Gummeson, K.~Jamieson, and W.~Hu, ``{LAVA}:
  fine-grained {3D} indoor wireless coverage for small {IoT} devices,'' in
  \emph{Proc. ACM SIGCOMM}, 2021, pp. 123--136.

\bibitem{shen2018fractional}
K.~Shen and W.~Yu, ``Fractional programming for communication systems—part
  {I}: Power control and beamforming,'' \emph{IEEE Trans. Signal Process.},
  vol.~66, no.~10, pp. 2616--2630, 2018.

\bibitem{dinkelbach1967nonlinear}
W.~Dinkelbach, ``On nonlinear fractional programming,'' \emph{Manage. Sci.},
  vol.~13, no.~7, pp. 492--498, 1967.

\bibitem{cvx}
M.~Grant and S.~Boyd, ``{CVX}: Matlab software for disciplined convex
  programming, version 2.1,'' \url{http://cvxr.com/cvx}, 2014.

\bibitem{spivak1994calculus}
M.~Spivak, \emph{Calculus}.\hskip 1em plus 0.5em minus 0.4em\relax Publish or
  Perish, 1994.

\bibitem{gander1989constrained}
W.~Gander, G.~H. Golub, and U.~Von~Matt, ``A constrained eigenvalue problem,''
  \emph{Linear Algebra and its applications}, vol. 114, pp. 815--839, 1989.

\bibitem{krantz2002implicit}
S.~G. Krantz and H.~R. Parks, \emph{The implicit function theorem: history,
  theory, and applications}.\hskip 1em plus 0.5em minus 0.4em\relax Springer
  Science \& Business Media, 2002.

\bibitem{park2017dynamic}
S.~Park, A.~Alkhateeb, and R.~W. Heath, ``Dynamic subarrays for hybrid
  precoding in wideband mmwave {MIMO} systems,'' \emph{IEEE Trans. Wireless
  Commun.}, vol.~16, no.~5, pp. 2907--2920, 2017.

\bibitem{yu2007transmitter}
W.~Yu and T.~Lan, ``Transmitter optimization for the multi-antenna downlink
  with per-antenna power constraints,'' \emph{IEEE Trans. Signal Process.},
  vol.~55, no.~6, pp. 2646--2660, 2007.

\bibitem{vu2011miso}
M.~Vu, ``{MISO} capacity with per-antenna power constraint,'' \emph{IEEE Trans.
  Commun.}, vol.~59, no.~5, pp. 1268--1274, 2011.

\bibitem{tasci2022new}
R.~A. Tasci, F.~Kilinc, E.~Basar, and G.~C. Alexandropoulos, ``A new {RIS}
  architecture with a single power amplifier: Energy efficiency and error
  performance analysis,'' \emph{IEEE Access}, vol.~10, pp. 44\,804--44\,815,
  2022.

\bibitem{liu2021active}
K.~Liu, Z.~Zhang, L.~Dai, S.~Xu, and F.~Yang, ``Active reconfigurable
  intelligent surface: Fully-connected or sub-connected?'' \emph{IEEE Commun.
  Lett.}, vol.~26, no.~1, pp. 167--171, 2021.

\bibitem{swindlehurst2022channel}
A.~L. Swindlehurst, G.~Zhou, R.~Liu, C.~Pan, and M.~Li, ``Channel estimation
  with reconfigurable intelligent surfaces—a general framework,'' \emph{Proc.
  IEEE}, vol. 110, no.~9, pp. 1312--1338, 2022.

\bibitem{ren2022configuring}
S.~Ren, K.~Shen, Y.~Zhang, X.~Li, X.~Chen, and Z.-Q. Luo, ``Configuring
  intelligent reflecting surface with performance guarantees: Blind
  beamforming,'' \emph{IEEE Trans. Wireless Commun.}, 2022.

\bibitem{shen2018fractional2}
K.~Shen and W.~Yu, ``Fractional programming for communication systems—part
  {II}: Uplink scheduling via matching,'' \emph{IEEE Trans. Signal Process.},
  vol.~66, no.~10, pp. 2631--2644, 2018.

\end{thebibliography}
